\documentclass[letterpaper,journal]{IEEEtran}
\IEEEoverridecommandlockouts
\usepackage{cite}
\usepackage{amsmath,amssymb,amsfonts}
\usepackage{algorithmic}
\usepackage{algorithm}
\usepackage{graphicx}
\usepackage{textcomp}
\usepackage{makecell}
\usepackage{xcolor}
\usepackage{booktabs}
\usepackage{pifont}
\usepackage{multirow}

\usepackage{titlesec}
\titlespacing*{\section}{0pt}{0.5\baselineskip}{0.3\baselineskip}
\titlespacing*{\subsection}{0pt}{0.4\baselineskip}{0.2\baselineskip}
\titlespacing*{\subsubsection}{0pt}{0.3\baselineskip}{0.1\baselineskip}

\usepackage[font=footnotesize]{caption}

\usepackage{enumitem}
\setlist{nosep}

\newcommand{\partialmark}{$\triangle$} 
\def\BibTeX{{\rm B\kern-.05em{\sc i\kern-.025em b}\kern-.08em
    T\kern-.1667em\lower.7ex\hbox{E}\kern-.125emX}}

\begin{document}

\title{A Cascaded Graph Neural Network for Joint Root Cause Localization and Analysis in Edge Computing Environments}


\author{Duneesha Fernando, 
Maria A. Rodriguez, 
Rajkumar Buyya
\thanks{D. Fernando, M. A. Rodriguez, and R. Buyya are with the Quantum Cloud Computing and Distributed Systems (qCLOUDS) Laboratory, School of Computing and Information
Systems, The University of Melbourne, Parkville, Australia.}
}

\maketitle

\begin{abstract}

Edge computing environments host increasingly complex microservice-based IoT applications that are prone to performance anomalies propagating across dependent services. Identifying the faulty component (root cause localization) and the underlying fault type (root cause analysis) is essential for timely mitigation. Supervised graph neural networks (GNNs) currently represent the state of the art for joint root cause localization and analysis. However, existing approaches rely on centralized processing over full-system graphs, leading to high inference latency and limited scalability in large, distributed edge environments. In this paper, we propose a cascaded GNN framework for joint RCL and fault type identification that explicitly addresses these scalability challenges. Our approach employs communication-driven clustering to partition large service graphs into highly interacting communities and a cascaded network with two subnetworks that perform hierarchical RCL/RCA. By restricting message passing to reduced and structured subgraphs, the proposed framework significantly lowers computational complexity while preserving critical dependency information. We evaluate the proposed method on the MicroCERCL benchmark and large-scale datasets generated using the iAnomaly simulation framework. Experimental results show that the cascaded architecture achieves diagnostic accuracy comparable to centralized GNN baselines while maintaining near-constant inference latency as graph size increases, enabling scalable and actionable AIOps in edge computing environments.
\end{abstract}

\begin{IEEEkeywords}
Edge computing, root cause localization, root cause analysis, anomaly detection, microservices, GNNs, performance diagnosis.
\end{IEEEkeywords}

\section{Introduction}
Edge-cloud integrated environments consist of geographically distributed devices with heterogeneous computing, storage, and networking capabilities. IoT applications are frequently architected as microservices and deployed in these distributed environments to satisfy the Quality of Service (QoS) requirements of each module while optimizing resource utilization \cite{al-doghman2023ai-enabled, WU2023Towards}. Over time, these microservice-based IoT applications are susceptible to performance anomalies caused by resource hogging (e.g., CPU or memory) and resource contention, which can negatively impact their QoS and violate their Service Level Agreements \cite{becker2020towards, Soualhia2019infrastructure}. Therefore, it is crucial to detect performance anomalies in microservice-based IoT applications in edge computing environments and ultimately mitigate them.

However, anomalies can propagate along communication and colocation dependencies, causing multiple services to appear anomalous despite a single underlying root cause \cite{scheinert2021arvalus, tian2023microgbpm}. 
Identifying the exact microservice responsible for an observed anomaly, known as Root Cause Localization (RCL), is essential for enabling timely and effective mitigation. Beyond locating the problematic component, it is equally important to determine why the anomaly occurred. For instance, the underlying cause may be CPU hogging, memory stress, or network delay. Determining the cause of the anomaly is referred to as Root Cause Analysis (RCA). All RCA studies perform RCL, while the analysis can be conducted at various levels \cite{wang2024comprehensivesurveyrootcause, xin2023causalrca}. Some RCA studies focus on identifying the specific metrics responsible for the anomaly \cite{wu2021microdiag, zhang2021aamr, zhu2024hemirca}, while others concentrate on determining the type of fault \cite{zhang2023diagfusion, zhu2024microirc, li2022dejavu}. Additionally, there is a more detailed level of RCA that examines issues at the source code level, which is valuable for developing long-term solutions to performance problems \cite{yu2023nezha}. In this work, we focus on the type of RCA that combines fault-type diagnosis together with RCL. This level of analysis provides essential information for autonomous AIOps (Artificial Intelligence for IT Operations) systems to address performance issues effectively \cite{li2022dejavu}.

Although RCL and RCA are often discussed separately, the two tasks are interconnected. They both stem from the same anomalous event and rely on similar signals such as temporal performance patterns, inter-service dependencies, and anomaly propagation behaviour. For this reason, recent studies advocate performing RCL and RCA jointly rather than in isolation \cite{li2022dejavu, zhu2024microirc, zhang2023diagfusion}. Supervised Graph Neural Networks (GNNs) currently represent the state of the art for joint RCL and RCA \cite{ren2023grace}. By learning RCL and RCA jointly, the model can share early feature extractors (such as CNN layers and GCN layers), allowing it to capture useful temporal patterns and inter-service dependencies that help both tasks. This enables faster and more accurate localization and diagnosis than training separate models \cite{zhang2023diagfusion, zhu2024microirc}.

Despite their effectiveness, GNNs have high computational complexity \cite{fu2025intelligentrclsurvey, scheinert2021arvalus}. In edge computing environments, which consist of a large number of devices and microservices, this translates to forming graphs with a large number of nodes as input for the GNN. The presence of such large-scale graphs amplifies the computational cost of message passing, resulting in slower diagnosis and delayed mitigation. 

To address this scalability challenge, we propose a cascaded GNN architecture designed for joint RCL and RCA, specifically designed for large-scale edge deployments. Our approach groups microservices into clusters based on their communication dependencies, with each cluster representing a highly interacting group of microservices. The cascaded GNN consists of two networks: the proposal network (P-Net) and the output network (O-Net). P-Net operates on individual clusters and produces a cluster-level output, which reduces the search space by focusing on a confined area. O-Net takes the latent representation of each cluster produced by P-Net and inter-cluster edges to create a graph representing the edge system, as input. By treating each cluster as a node, O-Net further minimizes its search space. As both networks work with reduced search spaces compared to a traditional GNN, the overall complexity of our cascaded network is lower. Although operating within a reduced receptive field may result in some information loss relative to a traditional GNN, our clustering technique effectively groups microservices based on communication dependencies, which mitigates this loss. This ensures that accuracy is maintained while improving efficiency. To the best of our knowledge, ours is the first work to introduce a computationally efficient cascaded GNN framework for joint RCL and RCA in edge computing environments, explicitly addressing the scalability and latency constraints present in large, distributed edge–cloud systems.  

We evaluated our proposed cascaded GNN approach on the publicly available MicroCERCL benchmark dataset \cite{zhu2024microcercl} as well as on large-scale datasets generated using the iAnomaly simulation framework \cite{fernando2024ianomaly}. Experimental results on MicroCERCL show that the cascaded model achieves diagnostic accuracy comparable to a centralized GNN baseline, while scalability experiments on iAnomaly demonstrate near-constant inference time, in contrast to the centralized model, whose latency increases with graph size.

The rest of the paper is organized as follows. Section \ref{sec:related_work} reviews the background and related work on root cause localization and analysis, joint RCL/RCA techniques, and efficiency-oriented diagnosis methods. Section \ref{sec:centralized_gnn} presents a centralized GNN architecture for joint RCL and RCA that serves as a baseline for comparison. Section \ref{sec:cascaded_gnn} introduces the proposed cascaded GNN framework, including the communication-driven clustering strategy and hierarchical inference design. Section \ref{sec:perf_eval} reports the experimental evaluation and scalability analysis on both real-world and large-scale simulated datasets. Finally, Section \ref{sec:conclusions} concludes the paper and outlines directions for future research.

\section{Background and Related Work}
\label{sec:related_work}

In this section, we provide an overview of the background and related work on root cause localization and analysis in microservice-based edge and cloud systems. We first review existing RCA approaches from the perspective of diagnostic granularity, ranging from coarse-grained localization to fine-grained fault type and code-level analysis. We then discuss recent techniques that perform RCL and RCA jointly, with a particular focus on supervised graph-based methods that represent the current state of the art. Finally, we examine prior efforts on improving the efficiency and scalability of RCA techniques, highlighting the limitations of existing solutions in large-scale edge environments and motivating the need for our cascaded GNN framework.

\subsection{Levels of Root Cause Analysis}

RCA in microservices-based systems spans a spectrum of diagnostic granularity, ranging from coarse-grained localization to fine-grained diagnosis. At the coarsest level, RCL, also referred to as coarse-grained localization, aims to identify the microservice responsible for an observed performance anomaly. RCL is generally considered as the first and most fundamental step of RCA, and is performed, either explicitly or implicitly, by almost all RCA approaches \cite{wang2024comprehensivesurveyrootcause, xin2023causalrca}. A number of studies focus exclusively on this task, including MicroRCA \cite{wu2020microrca}, MicroGBPM \cite{tian2023microgbpm}, MicroCERCL \cite{zhu2024microcercl}, and MicroEGRCL \cite{chen2022microegrcl}. While coarse-grained localization can often be achieved rapidly, identifying only the faulty microservice offers limited diagnostic value, as the underlying cause of the anomaly remains unknown, forcing Site Reliability Engineers to manually investigate issues such as memory exhaustion, CPU contention, or abnormal request patterns \cite{wu2020microrca-followup, zhu2024hemirca}. Moreover, mitigation actions at this level, such as restarting or redeploying an entire service, may fail to address the true root cause, introduce unnecessary disruption to dependent services, and prolong recovery times compared to fine-grained mitigation efforts \cite{xin2023causalrca}. 

Beyond microservice-level localization, RCA techniques increasingly incorporate finer-grained analysis. One intermediate level is culprit metric localization, which seeks to identify the specific performance metrics (e.g., CPU usage, latency, memory consumption) that contribute to the detected anomaly. Representative approaches in this category include MicroDiag \cite{wu2021microdiag}, AAMR \cite{zhang2021aamr}, HeMIRCA \cite{zhu2024hemirca}, CausalRCA \cite{xin2023causalrca}, PDiagnose \cite{hou2021pdiagnose}, and Wu et al. \cite{wu2020microrca-followup}. This level of analysis enables more targeted mitigation actions, such as resource scaling or throttling, which often result in faster recovery and lower system disruption than service-level restarts \cite{xin2023causalrca}. Both RCL-only approaches and those combining RCL with culprit metric localization are typically formulated as ranking problems, where microservices or metrics are ordered according to their likelihood of being the root cause.


A more advanced level of RCA focuses on fault type identification, which aims to determine the underlying cause of the anomaly itself. At this level, anomalies are attributed to specific fault categories such as CPU hogging, memory leaks, network delay, or I/O contention. Approaches including DiagFusion \cite{zhang2023diagfusion}, MicroIRC \cite{zhu2024microirc}, DejaVu \cite{li2022dejavu}, Brandon et al. \cite{brandon2020graph-based}, and MicroHECL \cite{liu2021microhecl} explicitly address this problem by classifying anomalies into predefined fault types. Fault type identification offers a key advantage over metric-level diagnosis: it directly maps observed anomalous behavior to semantically meaningful system faults, enabling more precise and automated mitigation actions. Consequently, this class of approaches is commonly formulated as a classification problem.

At the finest granularity, RCA operates at the source-code level, where the objective is to localize performance anomalies to specific code regions or components. Nezha \cite{yu2023nezha} exemplifies this category by identifying performance issues directly at the code level. While such approaches are valuable for developing long-term and permanent fixes, they often require extensive instrumentation of application source code or runtime systems, introducing significant overhead in production environments \cite{wu2020microrca-followup}. As a result, source-code-level RCA is generally less suitable for online diagnosis and real-time mitigation, particularly in latency-sensitive systems.

In this work, we focus on an RCA setting that combines fault type identification with RCL. This level of diagnosis provides essential information for autonomous AIOps systems to effectively address performance issues and aligns closely with the notion of actionable diagnosis \cite{li2022dejavu}. To be truly actionable, a fault localization solution must provide two critical pieces of information simultaneously: (1) where the failure occurs (the faulty component, via RCL), and (2) what kind of failure occurs (the fault type). Together, this combination is often referred to as a failure unit \cite{li2022dejavu}. Different fault types (e.g., memory leaks versus CPU exhaustion) require fundamentally different mitigation strategies. An approach that only localizes the faulty service cannot autonomously determine whether to scale resources, adjust scheduling policies, or apply memory management techniques. By jointly identifying both the faulty component and the fault type, RCA results become directly actionable, enabling AIOps systems to move beyond simple alerting and toward targeted, automated mitigation strategies \cite{soldani2022survey}.

\subsection{Joint RCL and RCA techniques}

Although RCL and RCA are often treated as separate tasks, they are interconnected.  The location of the fault and the nature of the fault both originate from the same anomalous event and are driven by shared system signals, including temporal performance patterns, inter-service dependencies, and anomaly propagation behavior. For this reason, approaches that combine RCL with fault type identification have gained increasing attention in recent years \cite{li2022dejavu, xin2023causalrca, zhu2024microirc, zhang2023diagfusion}. 

Among the three main categories of RCA techniques, 1) deep learning, 2) pattern recognition, and 3) causal inference, as identified by Wu et al. \cite{wu2021microdiag}, joint RCL and RCA has predominantly been explored using deep learning and pattern-based methods \cite{xin2023causalrca}. Deep learning approaches, in particular, have focused on supervised GNNs, which currently represent the state of the art for joint RCL and fault type identification \cite{ren2023grace}. Representative examples include DiagFusion \cite{zhang2023diagfusion}, which employs GNNs for joint diagnosis, MicroIRC \cite{zhu2024microirc}, which combines a personalized random walk algorithm with a GNN model, and DejaVu \cite{li2022dejavu}, which utilizes Graph Attention Networks (GATs). On the other hand, pattern-based approaches, such as the study by Brandon et al. \cite{brandon2020graph-based}, perform joint diagnosis by matching anomalous subgraphs against a predefined library of fault patterns.

Supervised GNNs are now the state of the art in joint RCL and RCA due to their ability to model complex inter-service dependencies. While some may argue that supervised methods struggle to identify previously unseen fault types, empirical evidence indicates that a significant proportion of failures in operational systems are recurring, with reports suggesting this figure can be as high as 74\% \cite{wang2024comprehensivesurveyrootcause}. Although supervised approaches typically require substantial labeled data and training effort, purely unsupervised methods often fail to achieve fully automated, end-to-end RCA \cite{zhu2024microcercl}. For these reasons, GNN-based supervised approaches can serve as a practical and effective starting point for automating AIOps, especially in edge environments.

\subsection{Efficient RCL and RCA techniques}

\begin{table*}[t]
\footnotesize
\centering
\caption{Summary of cloud and edge RCL/A techniques: accuracy vs efficiency focus}
\label{tab:rcl_summary}
\resizebox{\textwidth}{!}{%
\begin{tabular}{|c|c|c|c|c|c|c|}
\hline
\textbf{Work} & \textbf{\makecell{Cloud-only/ \\ Cloud-Edge}} & \textbf{GNN} & \textbf{RCL/RCA Type} & \textbf{\makecell{Accuracy \\Focus}} & \textbf{\makecell{Efficiency \\Focus}} & \textbf{Remarks} \\
\hline
\makecell{\cite{tian2023microgbpm}, \cite{lee2023eadro}, \cite{scheinert2021arvalus}, \cite{zhu2024hemirca},\\ \cite{wu2021microdiag}, \cite{xin2023causalrca}, \cite{wu2020microrca-followup}, \cite{zhu2024microirc}} 
& Cloud-only & Mixed 
& RCL; RCA (Metric / Fault Type) 
& \checkmark & \texttimes 
& Focused solely on accuracy \\
\hline
\makecell{\cite{chen2022microegrcl}, \cite{wu2020microrca}, \cite{zhang2021aamr}, \cite{ren2023grace}, \cite{brandon2020graph-based},\\ \cite{li2022dejavu}, \cite{zhang2023diagfusion}, \cite{yu2021microrank}, \cite{yu2023nezha}} 
& Cloud-only & Mixed 
& \makecell{RCL; RCA (Metric / Fault / \\Resource Type)} 
& \checkmark & $\triangle$ 
& \makecell{Claim faster inference; \\efficiency not a design priority} \\
\hline
\makecell{PDiagnose \cite{hou2021pdiagnose}, \\MicroHECL \cite{liu2021microhecl}} 
& Cloud-only & No 
& RCL; RCA (Metric / Fault Type) 
& $\triangle$ & \checkmark 
& \makecell{Focused on improving efficiency; \\may compromise accuracy} \\
\hline
MicroCERCL \cite{zhu2024microcercl}
& Cloud--Edge & Yes 
& RCL Only 
& \checkmark & \texttimes 
& \makecell{Accuracy focus;\\ long inference time reported} \\
\hline
Kalinagac et al. \cite{kalinagac2023liability}
& Cloud--Edge & No 
& RCA (Fault Type) 
& \checkmark & \texttimes 
& Focused solely on accuracy \\
\hline
\textbf{Our Work} 
& Cloud--Edge & Yes 
& Joint RCL + RCA (Fault Type) 
& \checkmark & \checkmark 
& \makecell{Cascaded GNN with clustering; \\balances accuracy and efficiency}\\
\hline
\end{tabular}%
}
{\raggedright
\footnotesize
\checkmark = Primary focus / addressed, \partialmark = Partially addressed / may compromise, \texttimes = Not addressed / ignored
\par}
\end{table*}

Despite their effectiveness, GNNs are known to incur non-trivial computational overhead, primarily due to the cost of iterative message passing over graph structures \cite{fu2025intelligentrclsurvey, scheinert2021arvalus}. In edge computing environments, which often consist of a large number of distributed devices, microservices, and their interactions, this results in graphs with a large number of nodes and edges being provided as input to the GNN. As the graph size increases, the computational cost of GNN inference grows proportionally with the number of nodes, edges, and message-passing layers, leading to longer diagnosis latency. In particular, large-scale graphs amplify the cost of neighborhood aggregation and feature propagation, which must be performed for each node across multiple GNN layers. This added complexity can slow down root cause identification and fault diagnosis, delaying mitigation actions. This delay is particularly problematic in latency-sensitive edge environments where timely response to anomalies is critical. 

To address this scalability challenge, we propose a cascaded GNN architecture for joint RCL and RCA, specifically targeting large-scale edge deployments. The framework consists of two sequential networks operating on clusters formed from highly interacting microservices based on communication dependencies. By applying both stages of the cascaded model to reduced and structured subgraphs, the proposed approach significantly limits the effective search space, thereby lowering computational complexity compared to conventional GNNs. Communication-based clustering preserves critical dependency information, minimizing information loss associated with reduced receptive fields and maintaining diagnostic accuracy. As a result, the proposed framework achieves an effective balance between efficiency and accuracy. 

A comparison of cloud and edge RCL/A techniques in terms of their focus on efficiency and accuracy is shown in Table~\ref{tab:rcl_summary}. The majority of studies on cloud RCL/A have primarily focused on improvements in accuracy, evaluating this single aspect \cite{tian2023microgbpm}, \cite{lee2023eadro}, \cite{scheinert2021arvalus}, \cite{zhu2024hemirca}, \cite{wu2021microdiag}, \cite{xin2023causalrca}, \cite{wu2020microrca-followup}, \cite{zhu2024microirc}. Some research has also addressed efficiency, particularly the localization times of their proposed methods. For instance, AAMR \cite{zhang2021aamr}, MicroEGRCL \cite{chen2022microegrcl}, and Grace \cite{ren2023grace} claim to provide faster inference times. However, none of these studies has specifically designed their approaches with efficiency as a priority. 
In addition, existing edge-oriented solutions, including MicroCERCL \cite{zhu2024microcercl} and the RCA framework proposed by Kalinagac et al. \cite{kalinagac2023liability}, do not explicitly incorporate efficiency considerations. In particular, MicroCERCL reports longer inference times than unsupervised heuristic methods, largely due to the complexity of the network \cite{zhu2024microcercl}.

MicroHECL \cite{liu2021microhecl} and PDiagnose \cite{hou2021pdiagnose} are cloud RCL/A techniques that specifically aim to improve efficiency, particularly by providing faster localization speeds. MicroHECL achieves efficiency by efficiently traversing the service dependency graph and using pruning techniques to eliminate irrelevant service calls during anomaly propagation chain analysis, which further enhances efficiency. 
On the other hand, PDiagnose tries to reach efficiency by removing the computationally heavy dependency graph-building phase and utilizing a vote-based localization process 
on an anomaly queue. 
However, both approaches adopt relatively simplistic strategies that may compromise diagnostic accuracy. Given the need to strike an appropriate balance between effectiveness and efficiency, our proposed method leverages a GNN-based approach to reduce diagnosis time while maintaining high accuracy.

Overall, existing RCA techniques have made substantial progress in improving diagnostic accuracy, particularly through supervised GNN-based joint RCL and RCA models. However, most of these approaches are designed under a centralized processing assumption and operate on full-system graphs, which leads to high computational overhead in large-scale edge environments. To better understand the strengths and limitations of such centralized designs, we first review a representative centralized joint RCL/RCA GNN architecture in the next section. This analysis provides the foundation for introducing our proposed cascaded framework, which explicitly addresses the scalability and latency challenges identified in prior works.

\section{Centralized Joint RCL/RCA GNN Architecture}
\label{sec:centralized_gnn}

In this section, we present a representative centralized GNN architecture for joint RCL and fault type identification, which serves as a baseline for subsequent comparison with our proposed cascaded framework. We first describe the graph formulation used to model edge computing environments and their microservice dependencies (Section \ref{subsec:edge_graph_form}), followed by the construction of GNN input representations from node- and edge-level time-series metrics (Section \ref{subsec:gnn_input_rep}). We then review common design principles of centralized GNN-based joint RCL/RCA models (Section \ref{subsec:gnn_bg}). Finally, we introduce the overall centralized architecture adopted in this work (Section \ref{subsec:final_centr_archi}).

\subsection{Graph formulation for edge computing environments}
\label{subsec:edge_graph_form}
Let the edge infrastructure consist of a set of edge devices {\small\[
\mathcal{E} = \{ e_1, e_2, \dots, e_{|\mathcal{E}|} \},
\]}

and a set of microservices deployed within the environment
{\small\[
\mathcal{M} = \{ m_1, m_2, \dots, m_{|\mathcal{M}|} \}.
\]}

Each microservice \( m_i \in \mathcal{M} \) is deployed on exactly one edge device. The deployment mapping can be defined as
{\small\[
\delta: \mathcal{M} \to \mathcal{E},
\]}
where \(\delta(m_i)\) gives the device on which microservice \(m_i\) is deployed.

GNN-based analysis requires the system to be represented as a graph, where nodes capture computational entities and edges reflect structural or runtime relationships. In the context of edge-cloud microservice systems, we model the environment using a topology graph
{\small\[
\mathcal{G} = (\mathcal{V}, \mathcal{L}),
\]}

where the set of vertices \(\mathcal{V}\) consists of both microservices and the edge devices on which they are deployed:
{\small\[
\mathcal{V} = \mathcal{M} \cup \mathcal{E}
\]}

The set of edges \(\mathcal{L}\) represents relationships between these vertices and is composed of two distinct types:
{\small\[
\mathcal{L} = \mathcal{L}_{\mathrm{comm}} \cup \mathcal{L}_{\mathrm{dep}}
\]}

\begin{enumerate}
    \item Communication edges (\(L_{comm}\)) capture runtime interactions among microservices, derived from trace data:
    \[
    \mathcal{L}_{\mathrm{comm}} \subseteq \mathcal{M} \times \mathcal{M},
    \]
    where \((m_i, m_j) \in \mathcal{L}_{\mathrm{comm}}\) if microservice \(m_i\) communicates with \(m_j\).
    \item Deployment edges (\(L_{dep}\)) link each microservice to the edge device on which it is hosted, obtained from the deployment configuration:
    \[
    \mathcal{L}_{\mathrm{dep}} = \{ (m_i, \delta(m_i)) \mid m_i \in \mathcal{M} \}.
    \]
\end{enumerate}

This unified heterogeneous graph structure enables joint reasoning over system execution behavior and physical deployment constraints.

Both nodes and edges are enriched with time-series features derived from monitoring metrics. Microservice nodes incorporate both resource-level metrics (e.g., CPU utilization, memory usage, disk I/O) and application-level metrics reflecting service behavior, whereas edge-device nodes are represented solely by resource-level metrics that describe device-level resource consumption. Edge features capture communication behavior (e.g., p50/p90/p99 latency percentiles), reflecting runtime dependencies and potential performance bottlenecks. 

\subsection{GNN input representation}
\label{subsec:gnn_input_rep}

From a modeling perspective, a GNN operates on three fundamental components: node features, graph connectivity, and (optionally) edge features. Let the node feature tensor be defined as
{\small\[
X \in \mathbb{R}^{|\mathcal{V}| \times F \times T},
\]}
where each node is represented by \(F\) feature dimensions observed over \(T\) timesteps. Since the graph contains \(|\mathcal{V}|\) nodes, the dimensions of the resulting feature tensor become \((|\mathcal{V}|\), \(F\), \(T)\).

The structural connectivity of the system is encoded using the edge index
{\small\[
\mathcal{L} \subseteq \mathcal{V} \times \mathcal{V},
\]}
capturing communication and deployment relationships among nodes.

Link-level properties and communication statistics are incorporated through the edge feature tensor, which is defined as
{\small\[
E \in \mathbb{R}^{|\mathcal{L}| \times F_e \times T},
\]}
where each edge is characterized by \(F_e\) feature dimensions measured over \(T\) timesteps. Given \(|\mathcal{L}|\) edges in the graph, the  dimensions of tensor \(E\) become \((|\mathcal{L}|\), \(F_e\), \(T)\). 

These components together constitute the input to the GNN model, 
enabling representation learning over both the structural and temporal characteristics 
of the topology graph.

\subsection{Background on centralized GNNs for joint RCL/RCA}
\label{subsec:gnn_bg}

When examining supervised GNN-based approaches employed by RCL and fault type identification studies, two dominant modeling paradigms can be observed. In the first paradigm, a single static graph is constructed, where temporal dependencies are learned first by extracting compact representations from time-series data, followed by spatial dependency learning using a GNN \cite{zhang2023diagfusion, zhu2024microirc, li2022dejavu}. In the second paradigm, a sequence of graphs is constructed (typically one graph per timestamp or time window) where spatial features are learned independently for each graph, and temporal dependencies are subsequently modeled across the graph sequence using temporal aggregation or sequence models \cite{zhu2024microcercl, ren2023grace}.

In the first category of approaches, temporal feature extraction is commonly performed using techniques such as LSTMs, CNNs or other sequence encoders
\cite{li2022dejavu, zhu2024microcercl, lee2023eadro}. The learned temporal embeddings are then used as node- and/or edge-level features for graph-based learning. We adopt this first paradigm due to its advantages in terms of computational efficiency, model scalability, and clear separation of temporal and structural learning, which is particularly beneficial for large-scale microservice systems \cite{li2022dejavu, zhang2023diagfusion}.

Following temporal feature extraction and message passing through the GNN layers, task-specific prediction heads are applied to support both RCL and fault type identification (RCA). These tasks differ in granularity and objective: RCL aims to identify the specific microservice responsible for the anomaly, whereas RCA determines the underlying anomaly type or fault category.

Within GNN-based frameworks, the RCL problem is typically formulated as a node-level ranking task \cite{zhang2023diagfusion, zhu2024microcercl, lee2023eadro}. The model produces a score for each microservice node, which is transformed into a series of probability values over all nodes (e.g., via a softmax activation), allowing the most likely faulty service to be ranked highest. In contrast, fault type identification is commonly framed as a graph-level multi-class classification task, where a graph-level readout layer aggregates learned node representations (e.g., using mean/max pooling or attention-based pooling), followed by a classification head that predicts the probability of each fault type \cite{zhang2023diagfusion, zhu2024microirc}. 

Following this design, two output heads are employed: a node-level prediction head for RCL and a graph-level prediction head for RCA, which can be trained either independently or jointly \cite{zhang2023diagfusion, lee2023eadro}. While some studies unify these outputs into a single prediction task \cite{li2022dejavu, zhu2024microirc}, we adopt a dual-head formulation. 

\begin{figure}[t]
    \centering
    \includegraphics[width=\columnwidth]{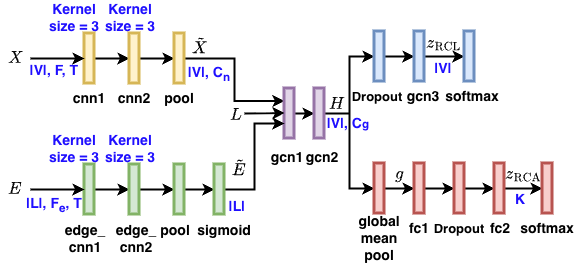}
    \caption{Centralised GNN architecture for joint RCL--RCA}
    \label{centralised_gnn}
\end{figure}

\subsection{Overview of centralised joint RCL/RCA GNN architecture}
\label{subsec:final_centr_archi}

Our baseline GNN architecture, as shown in Figure \ref{centralised_gnn}, follows the first modeling paradigm described in subsection \ref{subsec:gnn_bg}, where temporal patterns are learned prior to graph-based spatial reasoning. The architecture is composed of three sequential stages: (1) temporal feature extraction from nodes and edges, (2) spatial dependency learning using GCN-based message passing, and (3) two task-specific prediction heads for RCL and RCA. The early layers (temporal encoders and structural graph encoders) are shared across both tasks, enabling the network to jointly leverage temporal metric evolution and microservice dependency structure.

\subsubsection{Stage 1 - Temporal Feature Extraction}

To obtain compact representations from node- and edge-level time-series metrics, we employ separate 1D-CNN based temporal encoders, consistent with the feature processing approach discussed previously.

\textbf{(a) Node feature encoder.}
The input node tensor \(X \in \mathbb{R}^{|\mathcal{V}| \times F \times T}\) is passed through two CNN layers, followed by pooling across the temporal dimension:
{\begin{equation}
    X \xrightarrow{\text{CNN}_1} 
    \xrightarrow{\text{CNN}_2} 
    \xrightarrow{\text{Temporal Pool}} 
    \tilde{X} \in \mathbb{R}^{|\mathcal{V}| \times C_n},
\end{equation}}
where \(C_n\) is the output embedding dimension after temporal pooling. The pooling operation aggregates temporal information into a fixed-length representation, allowing downstream GNN layers to operate on time-compressed embeddings rather than raw sequences.

\textbf{(b) Edge feature encoder.}
Similarly, the edge attribute tensor \(E \in \mathbb{R}^{|\mathcal{L}| \times F_e \times T}\) is processed using two CNN layers, followed by pooling across the temporal dimension:
{\begin{equation}
    E \xrightarrow{\text{CNN}_1} 
    \xrightarrow{\text{CNN}_2} 
    \xrightarrow{\text{Temporal Pool}} 
    \tilde{E} \in \mathbb{R}^{|\mathcal{L}|}.
\end{equation}}
A sigmoid function is then applied to reduce the edge weights into a singular value in the range of \([0,1]\), facilitating the learning of necessary information for subsequent message passing stages. 

These embeddings form the initial node and edge features for the GNN layers.

\subsubsection{Stage 2 - Spatial Graph Learning}

The temporally encoded embeddings are passed through two stacked GCN layers to learn spatial dependencies across the topology graph:
{\begin{equation}
    (\tilde{X}, \mathcal{L}, \tilde{E}) \xrightarrow{\text{GCN}_1}
    \xrightarrow{\text{GCN}_2} 
    H \in \mathbb{R}^{|\mathcal{V}| \times C_g},
\end{equation}}
where \(C_g\) is the hidden dimension of the resulting structural representation. The edge embeddings \(\tilde{E}\) are used as edge weights, enabling the GCN to incorporate communication intensity or latency sensitivity when propagating information across the service graph.

These GCN layers act as structural encoders, shared by both downstream tasks. This shared representation is critical because both localization and fault type inference benefit from understanding inter-service influence patterns and anomaly propagation dynamics.

\subsubsection{Stage 3 - Task-Specific Prediction Heads}

After spatial encoding, the network branches into two independent output heads.

\textbf{(a) RCL Head – Node-level ranking.}
Designed to identify the faulty microservice, this branch operates directly on node embeddings:
{\begin{equation}
    H \xrightarrow{\text{Dropout}}
    \xrightarrow{\text{GCN}_3} 
    z_{\text{RCL}} \in \mathbb{R}^{|\mathcal{V}|},
\end{equation}}
\(z_{\text{RCL}}\) is the vector of logit values corresponding to all nodes in the graph. A softmax activation converts the logit values into probabilities, identified as \(\hat{p}_{\text{RCL}}\).

\textbf{(b) RCA Head – Graph-level multi-class classification.}
Here, node embeddings are aggregated via global mean pooling to obtain a graph-level representation \(g\), which is then processed through two fully connected layers with dropout to generate class logits:
{\begin{equation}
    H \xrightarrow{\text{GlobalMeanPool}} g
    \xrightarrow{\text{FC}_1}
    \xrightarrow{\text{Dropout}}
    \xrightarrow{\text{FC}_2}
    z_{\text{RCA}} \in \mathbb{R}^{K},
\end{equation}}
where \(K\) is the number of anomaly/fault types. A softmax activation converts the logit values into probabilities required for fault diagnosis, identified as \(\hat{p}_{\text{RCA}}\).


For all hidden GCN and CNN layers explained previously, we employ Exponential Linear Unit (ELU) as the activation function.

\subsubsection{Loss Formulation}
\label{subsubsec:loss_eq}
Subsequently, both prediction heads are optimized simultaneously using a joint loss function,
{\begin{equation}
\mathcal{L}_{\text{total}} = 
\lambda_{\text{RCL}} \cdot \mathcal{L}_{\text{RCL}} + 
\lambda_{\text{RCA}} \cdot \mathcal{L}_{\text{RCA}},
\end{equation}}

where $\mathcal{L}_{\text{RCL}}$ and $\mathcal{L}_{\text{RCA}}$ denote node-level and graph-level negative log-likelihood losses, for RCL and RCA, respectively, and $\lambda_{\text{RCL}}$ and $\lambda_{\text{RCA}}$ are weighting coefficients. 

Let $\hat{y}_{\text{RCL}}$ denote the one-hot encoded vector indicating the ground-truth root cause node. RCL loss for a single graph is defined as 

{
\begin{equation}
\mathcal{L}_{\text{RCL}} 
= - \frac{1}{|\mathcal{V}|}{\sum_{i=1}^{|\mathcal{V}|}}{y_{{RCL}}^i}\log(p^i_{RCL}),
\end{equation}
}

where \(y_{{RCL}}^i\) is the \(i^{th}\) element of $\hat{y}_{\text{RCL}}$, and \(p^i_{RCL}\) denotes the predicted probability of node \(i\) being the root cause, and $|\mathcal{V}|$ is the number of nodes in the graph.

Similarly, for the RCA task, let $\hat{y}_{\text{RCA}}$ denote the one-hot encoded vector indicating the fault type over $K$ fault classes. RCA loss for a single graph is defined as 

{
\begin{equation}
\mathcal{L}_{\text{RCA}} 
= - \frac{1}{|\mathcal{K}|}{\sum_{j=1}^{|\mathcal{K}|}}{y_{{RCA}}^j}\log(p^j_{RCA}),
\end{equation}
}

where \(y_{{RCA}}^j\) is the \(j^{th}\) element of $\hat{y}_{\text{RCA}}$, and \(p^j_{RCA}\) denotes the predicted probability of fault type \(j\) being the root cause.

Unless stated otherwise, we set $\lambda_{\text{RCL}} = \lambda_{\text{RCA}} = 0.5$ in our implementation. 

\section{Cascaded Joint RCL/RCA GNN Architecture}
\label{sec:cascaded_gnn}

While the baseline centralized GNN model is effective, it does not scale efficiently for large microservice deployments. The computational cost of each GCN layer is $\mathcal{O}(N^2)$, where $N=|\mathcal{V}|$ denotes the number of nodes in the graph \cite{fu2025intelligentrclsurvey, scheinert2021arvalus}. Since multiple such layers are employed to perform global message passing, the overall computational overhead is further amplified. As the microservice ecosystem expands, every layer must process all nodes and their connections, leading to rapidly increasing diagnosis latency. In edge--cloud environments with hundreds or thousands of distributed services, this quadratic complexity becomes a critical bottleneck, making centralized inference computationally expensive and impractical for real-time diagnosis.

To overcome this challenge, we propose a cascaded GNN architecture designed for joint RCL and RCA, specifically targeting large-scale edge deployments. Our method leverages communication-driven clustering, grouping microservices into clusters based on their interaction intensity. Such clustering serves two purposes: (i) it reduces graph size for the initial inference stage, and (ii) it aligns with real-world anomaly propagation patterns, as performance anomalies typically spread along communication paths. As a result, when an anomaly occurs, its effects are more likely to be confined within the boundaries of the identified cluster, allowing our architecture to diagnose root causes more efficiently without compromising accuracy.

In the following, we first describe our communication-driven clustering strategy for partitioning large service graphs into highly interacting communities (Section \ref{comm-cluster}). We then present the overall cascaded GNN architecture, including the design of the proposal network and output network (Section \ref{subsec:cascaded_archi}), followed by details of their joint optimization. Finally, we analyze the computational efficiency of the proposed framework and compare it with the centralized baseline (Section \ref{subsec:comp_efficiency_analysis}).

\subsection{Communication-driven clustering}

\label{comm-cluster}

\begin{algorithm}[t]
\small
\caption{Communication-driven clustering algorithm}
\label{alg:comm-cluster}
\begin{algorithmic}[1]
\STATE \textbf{Input}: System graph $\mathcal{G} = (\mathcal{V}, \mathcal{L})$, deployment mapping $\delta: \mathcal{M} \to \mathcal{E}$
\STATE \textbf{Output}: Cluster assignments $\phi(m)$ for microservices and $\psi(e)$ for edge devices

\STATE Construct communication subgraph $\mathcal{G}_{\mathrm{comm}} = (\mathcal{M}, \mathcal{L}_{\mathrm{comm}})$
\STATE Apply Louvain modularity optimization to obtain microservice communities $\{c_1,c_2,\dots,c_{|\mathcal{C}|}\}$
\FOR{each community $c_i$}
    \FOR{each microservice $m$ in $c_i$}
        \STATE Assign $\phi(m) = c_i$ 
    \ENDFOR
\ENDFOR

\STATE \textit{/* Edge-device cluster assignment */}
\FOR{each edge device $e \in \mathcal{E}$}
    \STATE Let $\mathcal{M}_e = \{m \in \mathcal{M}\mid \delta(m)=e\}$ \COMMENT{microservices deployed on $e$}
    \STATE Determine dominant microservice cluster \\ 
    \hspace{0.5cm}$c^* = \arg\max\limits_{c_i} |\{m \in \mathcal{M}_e \mid \phi(m)=c_i\}|$
    \STATE Assign $\psi(e) = c^*$
\ENDFOR
\STATE \textbf{Return}: $\phi(m)$ for all $m \in \mathcal{M}$, $\psi(e)$ for all $e \in \mathcal{E}$
\end{algorithmic}
\end{algorithm}

For cluster formation, we adopt the Louvain community detection algorithm \cite{blondel2008louvain}, a modularity-optimizing method well suited for large-scale networks. Algorithm \ref{alg:comm-cluster} presents our communication-driven clustering algorithm. Given the system topology graph \( \mathcal{G} =(\mathcal{V},\mathcal{L}) \), where \(\mathcal{V} = \mathcal{M} \cup \mathcal{E}\) consists of microservices and edge devices, and 
\(\mathcal{L} = \mathcal{L}_{\mathrm{comm}} \cup \mathcal{L}_{\mathrm{dep}}\) contains both communication and deployment relationships, we derive a communication-only subgraph for clustering. Specifically, we construct

{\small\[
\mathcal{G}_{\mathrm{comm}} = (\mathcal{M}, \mathcal{L}_{\mathrm{comm}}),
\]}

i.e., we extract a subgraph containing only microservice nodes and their communication edges, since anomaly propagation primarily follows runtime call paths. Louvain clustering is applied to \(\mathcal{G}_{\mathrm{comm}}\) to obtain a node-to-cluster assignment:

{\small\[
\phi: \mathcal{M} \to \mathcal\{c_1,c_2,\dots\,c_{|\mathcal{C}|}\},
\]}

where \(|\mathcal{C}|\) is the number of detected clusters. 

To extend clustering to devices, we assign each edge device to the cluster that contains the majority of its deployed microservices. Formally, for a device \(e\), let \(\mathcal{M}_e = \{ m \in \mathcal{M} \mid \delta(m) = e \},\)

We select the dominant community

{\small\[
\psi(e) = \arg\max_{c_i} \left|\{\, m \in \mathcal{M}_e \mid \phi(m) = c_i \,\}\right|.
\]}

resulting in a consistent cluster structure across both microservices and edge infrastructure. These communication-driven clusters serve as the foundation of our cascaded GNN pipeline.

\subsection{Overview of cascaded joint RCL/RCA GNN architecture}
\label{subsec:cascaded_archi}

\begin{figure*}[t]
    \centering
    \includegraphics[width=0.8\textwidth]{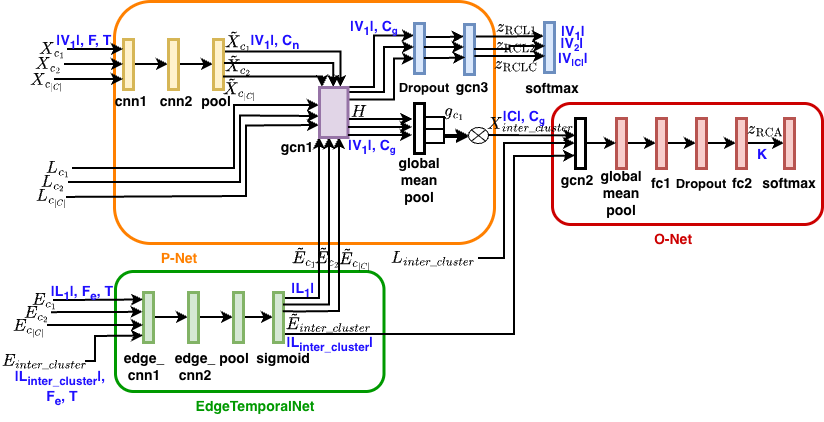}
    \caption{Cascaded GNN architecture for joint RCL--RCA}
    \label{cascadeded_gnn}
\end{figure*}

As shown in Figure \ref{cascadeded_gnn}, the cascaded GNN is composed of two main components: the proposal network (P-Net) and the output network (O-Net). We divided the baseline GNN into two networks: P-Net, which operates on individual clusters and generates cluster-level outputs, and O-Net, which considers each cluster as a node and operates over the graph of clusters and produces graph-level outputs by considering inter-cluster connectivity. In addition, the EdgeTemporalNet module, which consists of the edge feature encoder component of the baseline GNN architecture, is represented as an independent component since it is shared by both P-Net and O-Net for temporal edge feature embedding. The primary motivation behind this cascading design is to reduce the problem search space handled by each network, thereby lowering the computational burden on the GCN layers within those networks. This strategy aims to decrease overall diagnosis latency while still maintaining a high level of accuracy.

\subsubsection{Proposal Network (P-Net)}

To reduce the search space and improve computational efficiency, P-Net operates on individual clusters. Each cluster represents a highly communicating group of microservices as explained in Subsection \ref{comm-cluster}. The goal of P-Net is to learn intra-cluster dependencies and produce cluster-level outputs. It produces two such cluster-level outputs. The first is a cluster-level embedding that carries forward any necessary intra-cluster information for RCA. Embeddings from multiple such clusters are used as inputs in the subsequent O-Net. The second output is a vector of logit values corresponding to the nodes within the cluster. Since RCL is formulated as a node-level ranking problem and most anomaly propagation is confined to clusters (i.e., all the information required for RCL is typically available within a cluster), it is sufficient to perform RCL at the cluster level. Hence, P-Net is assigned the task of performing RCL. 

The inputs to P-Net are derived from the full graph attributes \(X\),\(\mathcal{L}\), and \(E\), but restricted per cluster. Let \(\mathcal\{c_1,c_2,\dots\,c_{|\mathcal{C}|}\}\) denote the set of clusters obtained during the clustering stage. For each cluster \(c_i\) in this set, we construct a corresponding subgraph with node features \(X_{c_i}\), edge index \(\mathcal{L}_{c_i}\), and edge features \(E_{c_i}\), containing only the nodes and intra-cluster edges belonging to \(c_i\). These cluster-specific inputs are then processed independently by P-Net. As illustrated in Figure \ref{cascadeded_gnn}, P-Net reuses the complete node feature encoder from the Temporal Feature Extraction stage of the baseline GNN. It also shares the EdgeTemporalNet-based edge feature encoder with O-Net as described earlier. The spatial graph learning component has a single GCN layer. The RCL prediction branch of the baseline model is retained up to the GCN layer, after which the cluster-level outputs are aggregated, and a global softmax is applied over all nodes (across all clusters) to produce the final RCL scores. Alongside node-level outputs for RCL, P-Net also produces the previously explained cluster-level embedding by applying global mean pooling to the GCN outputs (at its Spatial Graph Learning stage) within each cluster.

\subsubsection{Output Network (O-Net)}

O-Net treats each cluster as a node and the connectivity between clusters as edges, producing a graph-level output. The output embeddings of P-Net form the node features, while edge features are formed by aggregating the information from inter-cluster edges. Since RCA is formulated as a graph-level classification problem, O-Net is responsible for generating the final RCA prediction for the overall system. 

For O-Net, the node features \(X_{inter\_cluster}\) are the embeddings from P-Net, while the edge index \(\mathcal{L}_{inter\_cluster}\) is constructed from inter-cluster edges (i.e., edges connecting nodes from different clusters), with indices now mapped to cluster IDs rather than node IDs. The corresponding edge features \(E_{inter\_cluster}\) are first processed by the EdgeTemporalNet to extract compact time-series representations. When multiple edges exist between a pair of clusters, their encoded features are aggregated using global mean pooling to form a single inter-cluster edge representation, which is then passed to O-Net. As shown in Figure \ref{cascadeded_gnn}, O-Net consists of the architecture branch of the baseline model leading to the RCA head, and outputs class probabilities corresponding to fault types. 

\subsubsection{Joint Task Optimization}

Similar to the baseline architecture, the RCL and RCA outputs of the cascaded model are also optimized jointly using the same multi-task loss function mentioned in Section \ref{subsubsec:loss_eq}, enabling both objectives to be learned simultaneously while allowing shared representations to benefit both tasks.

\subsection{Computational efficiency analysis}
\label{subsec:comp_efficiency_analysis}

To assess how effectively the cascaded design meets its objective, we can analyze its computational behavior relative to the baseline GNN. Although the proposed model also utilizes a total of three GCN layers (two in P-Net and one in O-Net, equivalent to the baseline), the key efficiency gain arises from where each layer operates. Instead of processing all \(N\) nodes in a single graph, P-Net operates within each cluster. Let the set of nodes be partitioned into clusters of sizes \(n_1,n_2,\dots,n_k\) such that \(N = \sum_{i=1}^{k} n_i\). For a typical GCN layer with message passing complexity \(O(N^2)\), processing clusters independently yields a total complexity proportional to \(N = \sum_{i=1}^{k} n_i\) \cite{fu2025intelligentrclsurvey, scheinert2021arvalus}. Since

{\footnotesize
\[
N^2 = \left( \sum_{i=1}^{k} n_i \right)^2 
     = \sum_{i=1}^{k} n_i^2 + 2\sum_{i \neq j} n_i n_j.
\]}

and all \(n_i > 0\), it follows that:
{\footnotesize\[
N^2 > \sum_{i=1}^{k} n_i^2,
\]
}

meaning the work performed by P-Net is strictly lower than processing the entire graph at once. Moreover, clusters can be processed independently and thus parallelized, further lowering inference latency. O-Net then operates on a compact graph with \(|\mathcal{C}| \ll N\) nodes, leading to an additional drop in computation in the last GCN layer. 

Despite the compression introduced by clustering, accuracy is preserved because most anomaly propagation remains localized within clusters (thereby allowing P-Net to operate with minimal context loss) and O-Net recovers cross-cluster information through aggregated inter-cluster edges. Thus, the cascaded architecture can achieve the desired goals: reduced computational cost and faster diagnosis, while retaining accuracy, as validated in the next section.

\section{Performance Evaluation}
\label{sec:perf_eval}

In this section, we evaluate the proposed cascaded joint RCL/RCA GNN architecture in terms of both diagnostic effectiveness and scalability. We first describe the experimental setup, datasets, and implementation details for both MicroCERCL and iAnomaly (Section \ref{subsec:experimental_setup}). We then report accuracy and latency results on MicroCERCL, comparing the cascaded model against the centralized baseline under a medium-scale setting (Section \ref{subsec:microcercl_results}). Next, we analyze scalability on iAnomaly by varying graph size up to 10{,}000 nodes and measuring end-to-end inference time under different clustering configurations (Section \ref{subsec:scalability_analysis}). Finally, we conduct an ablation study to quantify the impact of joint learning, clustering strategy, and GNN layer type on performance (Section \ref{subsec:ablation_study}).

\subsection{Experimental setup and implementation details}
\label{subsec:experimental_setup}

We evaluated the proposed cascaded GNN approach using the publicly available MicroCERCL dataset\footnote{https://github.com/WDCloudEdge/MicroCERCL} \cite{zhu2024microcercl}. This dataset represents the first large-scale benchmark for cloud–edge collaborative microservice systems and remains the most comprehensive hybrid deployment dataset available to date. It contains data collected from 81 microservices belonging to four applications: SockShop, Hipster, Bookinfo, and the newly introduced AI-Edge. These microservices are deployed across four cloud servers and two groups of two edge servers, following a communication frequency-based application placement strategy.

To reflect realistic production environments, the dataset integrates a wide range of anomalies. Using ChaosMesh, the authors injected application-level anomalies, such as CPU resource exhaustion in containers, memory leaks, and network latency. Additionally, Linux kernel traffic control (TC) was employed to simulate kernel-level network failures—including packet loss, duplication, corruption, disorder, delay, and jitter—between cloud and edge nodes. These injections enable systematic evaluation under heterogeneous and complex failure conditions.

The dataset contains 682 fault scenarios, providing substantial diversity in anomaly types and intensities. For our experiments, we adopted a training, validation, and testing split of 60:20:20. For each scenario, distributed trace data were extracted to construct service dependency graphs, while corresponding system and application metrics were used for anomaly detection and as node and edge features for the GNN models. 

Although MicroCERCL represents a large-scale benchmark in terms of deployment realism and anomaly diversity, it can be regarded as a medium-scale dataset for efficiency-oriented evaluations. In particular, its graph sizes are insufficient to fully characterize the scalability behavior of RCL and RCA models under increasing system complexity. To evaluate the computational efficiency of the proposed approach under larger input sizes, we therefore employed a dataset generation tool based on the iAnomaly framework \cite{fernando2024ianomaly}. It enables the generation of service dependency graphs ranging from 50 to 10{,}000 nodes, thereby supporting systematic scalability analysis. Unlike purely synthetic simulators that rely on artificial workloads, the iAnomaly generator populates these graphs with anomalous traces derived from real execution data produced by the iAnomaly emulator. This ensures that the generated datasets retain realistic temporal and causal characteristics.

The underlying data originate from a diverse set of IoT applications, spanning lightweight sensor processing pipelines to computation-intensive services such as camera-based face detection and recognition. These applications are executed under a wide range of injected anomalies, including resource saturation, service failures, and network congestion. Consequently, the resulting datasets provide a realistic and scalable testbed for evaluating both the accuracy and efficiency of the proposed cascaded GNN framework.

Anomaly detection was performed on these datasets using the BIRCH clustering-based algorithm from Scikit-learn, following the configuration recommended by the MicroCERCL authors. Specifically, the anomaly sensitivity threshold \(\beta\) was set to 0.07, balancing anomaly detection accuracy and noise reduction.

Both the centralized and cascaded GNN models were implemented in Python 3.10 using PyTorch Geometric 2.6.1\footnote{https://pytorch-geometric.readthedocs.io/en/latest/}. Model hyperparameters were optimized using the Tree-structured Parzen Estimator (TPE)-based Bayesian optimization method \cite{bergstra2011tpe}. The proposed communication-driven clustering algorithm (Algorithm~\ref{alg:comm-cluster}) was implemented in Python. All experiments, including hyperparameter tuning, were conducted on the Spartan HPC cluster\footnote{https://dashboard.hpc.unimelb.edu.au/}. Model training was performed using the Adam optimizer \cite{kingma2017adamoptimization}.

\subsection{MicroCERCL evaluation results}
\label{subsec:microcercl_results}

Before comparing our proposed cascaded GNN approach against the centralized baseline GNN model, we first validate the accuracy of the centralized GNN baseline by benchmarking it against the results originally reported in the MicroCERCL paper. Notably, the MicroCERCL authors report only RCL performance for two applications (HH and SH) measured using top-$K$ localization accuracy. Their results are shown in Table \ref{microcercl_rcl_results}.

\begin{table}[h]
\footnotesize
\centering
\caption{RCL accuracy reported in the MicroCERCL paper}
\label{microcercl_rcl_results}
\begin{tabular}{lccccc}
\toprule
\textbf{Application} & \textbf{Acc@1} & \textbf{Acc@2} & \textbf{Acc@3} & \textbf{Acc@5} & \textbf{Acc@10} \\
\midrule
HH & 0.632 & 0.756 & 0.796 & 0.833 & 0.896 \\
SH & 0.607 & 0.732 & 0.792 & 0.849 & 0.907 \\
\bottomrule
\end{tabular}
\end{table}

To verify that our centralized baseline is competitive, we first train it to perform RCL only. As shown in Table~\ref{tab:rcl_only}, our model achieves significantly higher top-$K$ localization accuracy compared to the results reported in the MicroCERCL paper. These results confirm that our centralized baseline is competitive and suitable for a fair comparison with the proposed cascaded approach.

\begin{table}[h]
\footnotesize
\centering
\caption{RCL-only performance of our centralized GNN baseline}
\label{tab:rcl_only}
\begin{tabular}{lccccc}
\toprule
\textbf{} & \textbf{Acc@1} & \textbf{Acc@3} & \textbf{Acc@5} & \textbf{MAR} & \textbf{MRR} \\
\midrule
Centralized (RCL only) & 0.9203 & 0.9783 & 0.9928 & 1.1594 & 0.9534 \\
\bottomrule
\end{tabular}
\end{table}

We also evaluate the centralized baseline on RCA as an independent task, formulated as a fault type classification problem. The results are shown in Table \ref{tab:rca_only}.

\begin{table}[h]
\footnotesize
\centering
\caption{RCA-only performance of our centralized GNN baseline}
\label{tab:rca_only}
\begin{tabular}{lcccc}
\toprule
\textbf{} & \textbf{Accuracy} & \textbf{Precision} & \textbf{Recall} & \textbf{F1-score} \\
\midrule
Centralized (RCA only) & 0.8478 & 0.8667 & 0.8478 & 0.8512 \\
\bottomrule
\end{tabular}
\end{table}

We then train the baseline to jointly perform RCL and RCA using a shared representation and a multi-task objective. The joint learning results are summarized in Table~\ref{tab:joint}.  Compared to training separate models for RCL and RCA, joint optimization has increased both localization performance and fault classification accuracy. This indicates that shared representations allow complementary information to be exploited across tasks, leading to more effective end-to-end root cause diagnosis.

\begin{table*}[t]
\footnotesize
\centering
\caption{Joint RCL and RCA performance of our centralized baseline}
\label{tab:joint}
\begin{tabular}{lccccc|cccc}
\toprule
 & \multicolumn{5}{c|}{\textbf{RCL}} & \multicolumn{4}{c}{\textbf{RCA}} \\
\textbf{} & \textbf{Acc@1} & \textbf{Acc@3} & \textbf{Acc@5} & \textbf{MAR} & \textbf{MRR} & \textbf{Accuracy} & \textbf{Precision} & \textbf{Recall} & \textbf{F1-score} \\
\midrule
Joint RCL \& RCA & 0.9275 & 0.9855 & 1.0000 & 1.1377 & 0.9550 & 0.8696 & 0.8877 & 0.8696 & 0.8715 \\
\bottomrule
\end{tabular}
\end{table*}

We next evaluate our proposed cascaded GNN architecture for joint RCL and RCA and compare it against the centralized joint model. The results are summarized in Table~\ref{tab:cascaded_joint}. Overall, the cascaded approach achieves localization and diagnosis performance that is comparable to the centralized baseline
. This demonstrates that decomposing the global graph into communication-driven clusters and processing them via a cascaded network does not degrade diagnostic accuracy.

\begin{table*}[t]
\footnotesize
\centering
\caption{Joint RCL and RCA performance of the cascaded GNN}
\label{tab:cascaded_joint}
\begin{tabular}{lccccc|cccc}
\toprule
 & \multicolumn{5}{c|}{\textbf{RCL}} & \multicolumn{4}{c}{\textbf{RCA}} \\
\textbf{} & \textbf{Acc@1} & \textbf{Acc@3} & \textbf{Acc@5} & \textbf{MAR} & \textbf{MRR} & \textbf{Accuracy} & \textbf{Precision} & \textbf{Recall} & \textbf{F1-score} \\
\midrule
Cascaded RCL \& RCA & 0.9130 & 0.9420 & 0.9493 & 1.9058 & 0.9316 & 0.8623 & 0.8703 & 0.8623 & 0.8640 \\
\bottomrule
\end{tabular}
\end{table*}

To assess inference efficiency, we measure the end-to-end diagnosis time for both joint models. For the centralized joint GNN, the average and median diagnosis times are 17.779~ms and 17.707~ms, respectively. In comparison, the cascaded joint GNN incurs an average diagnosis time of 20.237~ms and a median time of 20.273~ms. These results indicate that, under the scale of the MicroCERCL dataset (approximately 50 nodes), the cascaded architecture does not yet exhibit clear latency improvements over the centralized baseline.

This observation is expected, as the MicroCERCL dataset is a medium-scale dataset in terms of efficiency evaluations and does not sufficiently stress the scalability limits of centralized graph processing. In such settings, the overhead introduced by clustering and multi-stage inference can offset potential computational savings. Consequently, while the cascaded GNN maintains comparable diagnostic accuracy, its efficiency benefits are not fully realized at this scale.

To further examine scalability, we conduct additional experiments on the iAnomaly dataset, which contains significantly larger service graphs. These experiments are designed to stress-test the centralized model and assess how the cascaded architecture behaves as graph size increases. We discuss these results next.

\subsection{Scalability analysis on the iAnomaly dataset}
\label{subsec:scalability_analysis}

Using the iAnomaly dataset, we vary the service graph size from 50 to 10{,}000 nodes to enable a systematic comparison between the centralized joint GNN and the cascaded joint GNN under increasing graph sizes.

We measure the average end-to-end diagnosis time for three configurations: (i) the centralized joint RCL/RCA model, (ii) the cascaded joint model with a fixed number of clusters (set to 10), and (iii) the cascaded joint model with an adaptive number of clusters, where the number of clusters is selected proportionally to the graph size (approximately one cluster per 20 nodes).

\begin{figure}[t]
    \centering
    \includegraphics[width=\columnwidth]{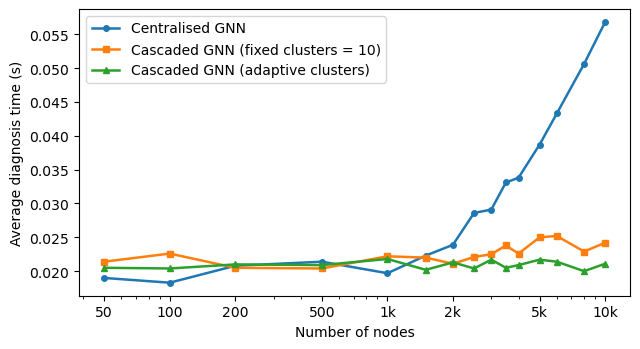}
    \caption{Scalability comparison on the iAnomaly dataset}
    \label{scalability_results}
\end{figure}

Figure~\ref{scalability_results} summarizes the results. As the number of nodes increases, the centralized model exhibits a clear upward trend in diagnosis time, indicating limited scalability due to global message passing over an increasingly large graph. The cascaded model with a fixed number of clusters reduces this growth but still shows a mild increase in latency as cluster sizes grow with graph scale. In contrast, when the number of clusters is adaptively controlled based on the graph size, the cascaded GNN maintains an approximately constant diagnosis time across all evaluated scales. 

These results confirm that the proposed cascaded GNN architecture is capable of preserving diagnostic accuracy while enabling scale-invariant inference, making it well-suited for large-scale microservice deployments in edge and cloud environments.

\subsection{Ablation study}
\label{subsec:ablation_study}

To analyze the contribution of individual design choices in the proposed cascaded GNN architecture, we conduct a series of ablation experiments focusing on joint learning, clustering strategy, and GNN layer type.

\begin{table*}[t]
\footnotesize
\centering
\caption{Ablation study on the proposed cascaded GNN architecture}
\label{tab:ablation_all}
\begin{tabular}{llccccc|cccc}
\toprule
\textbf{Category} & \textbf{Variant} 
& \multicolumn{5}{c|}{\textbf{RCL}} 
& \multicolumn{4}{c}{\textbf{RCA}} \\
\cmidrule(lr){3-7} \cmidrule(lr){8-11}
 &  & Acc@1 & Acc@3 & Acc@5 & MAR & MRR 
 & Acc. & Prec. & Rec. & F1 \\
\midrule

\multicolumn{2}{l}{\textbf{Proposed (Full Model)}} 
& \textbf{0.9130} & \textbf{0.9420} & \textbf{0.9493} & \textbf{1.9058} & \textbf{0.9316}
& \textbf{0.8623} & \textbf{0.8703} & \textbf{0.8623} & \textbf{0.8640} \\

\midrule

\multirow{2}{*}{Joint Learning}
& RCL only
& 0.8696 & 0.9420 & 0.9638 & 1.8116 & 0.9123
& -- & -- & -- & -- \\

& RCA only
& -- & -- & -- & -- & --
& 0.6739 & 0.6760 & 0.6739 & 0.6692 \\

\midrule

Clustering
& Random
& 0.7971 & 0.8623 & 0.8841 & 2.6884 & 0.8386
& 0.8478 & 0.8667 & 0.8478 & 0.8512 \\

\midrule

GNN Layer
& GAT
& 0.8478 & 0.9493 & 0.9855 & 1.3333 & 0.9067
& 0.8333 & 0.8422 & 0.8333 & 0.8346 \\

\bottomrule
\end{tabular}
\end{table*}

Table~\ref{tab:ablation_all} presents the results of the ablation study. The first row corresponds to the proposed full cascaded model with joint learning, communication-driven clustering, and GCN layers.

\paragraph{Impact of Joint Learning.}
Removing joint learning and training task-specific models leads to consistent performance degradation. For RCL, Acc@1 decreases from 0.9130 to 0.8696. For RCA, the impact is more pronounced: the F1-score drops sharply from 0.8640 to 0.6692. This substantial reduction indicates that shared representations across localization and diagnosis tasks improve feature generalization and enhance discriminative capability.

\paragraph{Impact of Clustering Strategy.}
Replacing communication-driven clustering with random clustering significantly degrades RCL performance. Acc@1 drops from 0.9130 to 0.7971, and MAR increases from 1.9058 to 2.6884, indicating poorer ranking quality. Although RCA F1-score decreases more modestly (from 0.8640 to 0.8512), the degradation across both tasks confirms that preserving communication dependencies during graph decomposition is critical for maintaining localization accuracy.

\paragraph{Impact of GNN Layer Type.}
Substituting GCN layers with GAT layers yields mixed results. While Acc@5 slightly increases (0.9493 to 0.9855) and MAR improves (1.9058 to 1.3333), Acc@1 decreases (0.9130 to 0.8478) and RCA F1-score drops from 0.8640 to 0.8346. Given that GAT introduces additional attention computation overhead, these results suggest that the added complexity does not translate into consistent performance gains. Overall, GCN layers offer a better balance between effectiveness and computational efficiency in the cascaded architecture.

\section{Conclusions and Future Work}
\label{sec:conclusions}

In this paper, we investigated the problem of joint root cause localization and fault type identification in large-scale edge–cloud microservice environments. While recent GNN-based approaches have demonstrated strong diagnostic accuracy, their reliance on centralized processing over full-system graphs leads to high inference latency and limited scalability in distributed edge settings. To address this challenge, we proposed a cascaded GNN architecture that decomposes the diagnosis task through communication-driven clustering and hierarchical graph reasoning.

Our framework consists of a proposal network that operates on individual communication-aware clusters to generate cluster-level outputs, followed by an output network which considers each cluster as a node and operates over the graph of clusters to produce graph-level outputs by considering inter-cluster connectivity. By restricting message passing to smaller, structurally coherent subgraphs and operating on a compact cluster-level graph, the proposed approach significantly reduces the effective search space while preserving critical dependency information.

Extensive evaluations on the MicroCERCL benchmark and large-scale datasets generated using iAnomaly demonstrate that the cascaded architecture achieves diagnostic accuracy comparable to centralized GNN models while providing substantial scalability benefits. In particular, experiments on large service graphs show that the proposed method maintains near-constant inference latency as system size increases, in contrast to the rapidly growing latency of centralized approaches. 

In future work, we plan to investigate distributed deployment strategies for the proposed cascaded framework. While the current design performs inference at a centralized location, deploying components of the model directly on edge devices could further reduce diagnosis latency by minimizing data transfer overhead. To support such decentralized operation, we will explore federated learning techniques for coordinating the proposal and output networks across distributed nodes. In addition, we aim to extend the current communication-driven clustering mechanism to jointly consider communication and colocation dependencies, enabling more deployment-aware partitioning in large-scale edge environments.

\bibliographystyle{IEEEtran}
	
{\footnotesize
\bibliography{reference}

@inproceedings{becker2020towards,
  author={Becker, Soeren and Schmidt, Florian and Gulenko, Anton and Acker, Alexander and Kao, Odej},
  booktitle={Proceedings of the 2020 IEEE International Conference on Big Data (Big Data)}, 
  title={Towards AIOps in Edge Computing Environments}, 
  publisher={IEEE},
  year={2020}, 
  pages={3470-3475},
  address = {Atlanta, GA, USA}}

@inproceedings{Soualhia2019infrastructure,
author = {Soualhia, Mbarka and Fu, Chunyan and Khomh, Foutse},
title = {Infrastructure Fault Detection and Prediction in Edge Cloud Environments},
year = {2019},
booktitle = {Proceedings of the 4th ACM/IEEE Symposium on Edge Computing},
address = {Arlington, Virginia},
series = {SEC '19},
pages = {222–235},
numpages = {14},
publisher = {Association for Computing Machinery},
}

@inproceedings{bergstra2011tpe,
 author = {Bergstra, James and Bardenet, R\'{e}mi and Bengio, Yoshua and K\'{e}gl, Bal\'{a}zs},
 booktitle = {Proceedings of the 25th Annual Conference on Advances in Neural Information Processing Systems},
 title = {Algorithms for Hyper-Parameter Optimization},
 address = {Granada, Spain},
 year = {2011},
 pages = {},
}

@ARTICLE{al-doghman2023ai-enabled,
  author={Al-Doghman, Firas and Moustafa, Nour and Khalil, Ibrahim and Sohrabi, Nasrin and Tari, Zahir and Zomaya, Albert Y.},
  journal={IEEE Transactions on Services Computing}, 
  title={AI-Enabled Secure Microservices in Edge Computing: Opportunities and Challenges}, 
  year={2023},
  volume={16},
pages={1485-1504},
}

@article{WU2023Towards,
title = {Towards cost-effective and robust AI microservice deployment in edge computing environments},
journal = {Future Generation Computer Systems},
volume = {141},
pages = {129-142},
year = {2023},
author = {Chunrong Wu and Qinglan Peng and Yunni Xia and Yong Jin and Zhentao Hu}
}

@inproceedings{fernando2024ianomaly,
author = {Duneesha Fernando and Maria A. Rodriguez and Rajkumar Buyya},
title = {iAnomaly: A Toolkit for Generating Performance Anomaly Datasets in Edge-Cloud Integrated Computing Environments},
year = {2024},
booktitle = {Proceedings of the 17th IEEE/ACM International Conference on Utility and Cloud Computing},
address = {Sharjah, UAE},
series = {UCC'24},
publisher = {Association for Computing Machinery},
pages={},
}

@article{soldani2022survey,
author = {Soldani, Jacopo and Brogi, Antonio},
title = {Anomaly Detection and Failure Root Cause Analysis in (Micro) Service-Based Cloud Applications: A Survey},
year = {2022},
volume = {55},
number = {3},
journal = {ACM Computing Surveys},
articleno = {59},
numpages = {39},
}

@misc{zhu2024microcercl,
      title={Root Cause Localization for Microservice Systems in Cloud-edge Collaborative Environments}, 
      author={Yuhan Zhu and Jian Wang and Bing Li and Xuxian Tang and Hao Li and Neng Zhang and Yuqi Zhao},
      year={2024},
      eprint={2406.13604},
      archivePrefix={arXiv},
      primaryClass={cs.SE},
      url={https://arxiv.org/abs/2406.13604}, 
}

@INPROCEEDINGS{scheinert2021arvalus,
  author={Scheinert, Dominik and Acker, Alexander and Thamsen, Lauritz and Geldenhuys, Morgan K. and Kao, Odej},
  booktitle={Proceedings of the 2021 IEEE/ACM International Workshop on Cloud Intelligence (CloudIntelligence)}, 
  title={Learning Dependencies in Distributed Cloud Applications to Identify and Localize Anomalies}, 
  year={2021},
  volume={},
  number={},
  pages={7-12},
  doi={10.1109/CloudIntelligence52565.2021.00011}}

@article{tian2023microgbpm,
author = {Tian, Wei and Zhang, Haitao and Yang, Neng and Zhang, Yepeng},
title = {Graph-Based Root Cause Localization in Microservice Systems with Protection Mechanisms},
journal = {International Journal of Software Engineering and Knowledge Engineering},
volume = {33},
number = {08},
pages = {1211-1238},
year = {2023}
}

@article{fu2025intelligentrclsurvey,
author = {Fu, Nan and Cheng, Guang and Teng, Yue and Dai, Guangye and Yu, Shui and Chen, Zihan},
title = {Intelligent Root Cause Localization in MicroService Systems: A Survey and New Perspectives},
year = {2025},
issue_date = {December 2025},
publisher = {Association for Computing Machinery},
address = {New York, NY, USA},
volume = {57},
number = {12},
issn = {0360-0300},
journal = {ACM Computing Surveys},
month = jul,
articleno = {325},
numpages = {37}
}

@misc{wang2024comprehensivesurveyrootcause,
      title={A Comprehensive Survey on Root Cause Analysis in (Micro) Services: Methodologies, Challenges, and Trends}, 
      author={Tingting Wang and Guilin Qi},
      year={2024},
      eprint={2408.00803},
      archivePrefix={arXiv},
      primaryClass={cs.SE},
      url={https://arxiv.org/abs/2408.00803}, 
}

@INPROCEEDINGS{kalinagac2023liability,
  author={Kalinagac, Onur and Soussi, Wissem and Anser, Yacine and Gaber, Chrystel and Gür, Gürkan},
  booktitle={Proceedings of the ICC 2023 - IEEE International Conference on Communications}, 
  title={Root Cause and Liability Analysis in the Microservices Architecture for Edge IoT Services}, 
  year={2023},
  volume={},
  number={},
  pages={3277-3283},
  doi={10.1109/ICC45041.2023.10279721}}

@inproceedings{liu2021microhecl,
author = {Liu, Dewei and He, Chuan and Peng, Xin and Lin, Fan and Zhang, Chenxi and Gong, Shengfang and Li, Ziang and Ou, Jiayu and Wu, Zheshun},
title = {MicroHECL: high-efficient root cause localization in large-scale microservice systems},
year = {2021},
isbn = {9780738146690},
publisher = {IEEE Press},
booktitle = {Proceedings of the 43rd International Conference on Software Engineering: Software Engineering in Practice},
pages = {338–347},
numpages = {10},
location = {Virtual Event, Spain},
series = {ICSE-SEIP '21}
}

@INPROCEEDINGS{wu2020microrca,
  author={Wu, Li and Tordsson, Johan and Elmroth, Erik and Kao, Odej},
  booktitle={Proceedings of the NOMS 2020 - 2020 IEEE/IFIP Network Operations and Management Symposium}, 
  title={MicroRCA: Root Cause Localization of Performance Issues in Microservices}, 
  year={2020},
  volume={},
  number={},
  pages={1-9},
  doi={10.1109/NOMS47738.2020.9110353}}

@INPROCEEDINGS{zhang2021aamr,
  author={Zhang, Zekun and Li, Bing  and Wang, Jian and Liu, Yongqiang},
  booktitle={Proceedings of the SEKE'21}, 
  title={AAMR: Automated Anomalous Microservice Ranking in Cloud-Native Environment}, 
  year={2021},
  volume={},
  number={},
  pages={86-91},
}

@article{xin2023causalrca,
author = {Xin, Ruyue and Chen, Peng and Zhao, Zhiming},
title = {CausalRCA: Causal inference based precise fine-grained root cause localization for microservice applications},
year = {2023},
issue_date = {Sep 2023},
publisher = {Elsevier Science Inc.},
address = {USA},
volume = {203},
number = {C},
issn = {0164-1212},
journal = {Journal of Systems and Software},
month = sep,
numpages = {13}
}

@article{zhu2024hemirca,
author = {Zhu, Zhouruixing and Lee, Cheryl and Tang, Xiaoying and He, Pinjia},
title = {HeMiRCA: Fine-Grained Root Cause Analysis for Microservices with Heterogeneous Data Sources},
year = {2024},
issue_date = {November 2024},
publisher = {Association for Computing Machinery},
address = {New York, NY, USA},
volume = {33},
number = {8},
issn = {1049-331X},
journal = {ACM Transactions on Software Engineering and Methodology},
month = nov,
articleno = {200},
numpages = {25}
}

@inproceedings{yu2023nezha,
author = {Yu, Guangba and Chen, Pengfei and Li, Yufeng and Chen, Hongyang and Li, Xiaoyun and Zheng, Zibin},
title = {Nezha: Interpretable Fine-Grained Root Causes Analysis for Microservices on Multi-modal Observability Data},
year = {2023},
isbn = {9798400703270},
booktitle = {Proceedings of the 31st ACM Joint European Software Engineering Conference and Symposium on the Foundations of Software Engineering},
pages = {553–565},
numpages = {13},
address = {San Francisco, CA, USA},
series = {ESEC/FSE 2023}
}

@INPROCEEDINGS{wu2021microdiag,
  author={Wu, Li and Tordsson, Johan and Bogatinovski, Jasmin and Elmroth, Erik and Kao, Odej},
  booktitle={Proceedings of the 2021 IEEE/ACM International Workshop on Cloud Intelligence (CloudIntelligence)}, 
  title={MicroDiag: Fine-grained Performance Diagnosis for Microservice Systems}, 
  year={2021},
  volume={},
  number={},
  pages={31-36},
  doi={10.1109/CloudIntelligence52565.2021.00015}}

@article{zhu2024microirc,
author = {Zhu, Yuhan and Wang, Jian and Li, Bing and Zhao, Yuqi and Zhang, Zekun and Xiong, Yiming and Chen, Shiping},
title = {MicroIRC: Instance-level Root Cause Localization for Microservice Systems},
year = {2024},
issue_date = {Oct 2024},
publisher = {Elsevier Science Inc.},
address = {USA},
volume = {216},
number = {C},
issn = {0164-1212},
journal = {Journal of Systems and Software},
month = oct,
numpages = {18}
}

@INPROCEEDINGS{ren2023grace,
  author={Ren, Rui and Wang, Yang and Liu, Fengrui and Li, Zhenyu and Tyson, Gareth and Miao, Tianhao and Xie, Gaogang},
  booktitle={Proceedings of the 2023 IEEE/ACM 31st International Symposium on Quality of Service (IWQoS)}, 
  title={Grace: Interpretable Root Cause Analysis by Graph Convolutional Network for Microservices}, 
  year={2023},
  volume={},
  number={},
  pages={1-4},
  doi={10.1109/IWQoS57198.2023.10188728}}

@INPROCEEDINGS{hou2021pdiagnose,
  author={Hou, Chuanjia and Jia, Tong and Wu, Yifan and Li, Ying and Han, Jing},
  booktitle={Proceedings of the 2021 IEEE International Conference on Parallel \& Distributed Processing with Applications, Big Data \& Cloud Computing, Sustainable Computing \& Communications, Social Computing \& Networking (ISPA/BDCloud/SocialCom/SustainCom)}, 
  title={Diagnosing Performance Issues in Microservices with Heterogeneous Data Source}, 
  year={2021},
  volume={},
  number={},
  pages={493-500},
  doi={10.1109/ISPA-BDCloud-SocialCom-SustainCom52081.2021.00074}}

@inproceedings{chen2022microegrcl,
author = {Chen, Ruibo and Ren, Jian and Wang, Lingfeng and Pu, Yanjun and Yang, Kaiyuan and Wu, Wenjun},
title = {MicroEGRCL: An Edge-Attention-Based Graph Neural Network Approach for Root Cause Localization in Microservice Systems},
year = {2022},
publisher = {Springer-Verlag},
address = {Berlin, Heidelberg},
doi = {10.1007/978-3-031-20984-0_18},
booktitle = {Service-Oriented Computing: 20th International Conference, ICSOC 2022, Seville, Spain, November 29 – December 2, 2022, Proceedings},
pages = {264–272},
numpages = {9},
location = {Seville, Spain}
}

@inproceedings{wu2020microrca-followup,
author = {Wu, Li and Bogatinovski, Jasmin and Nedelkoski, Sasho and Tordsson, Johan and Kao, Odej},
title = {Performance Diagnosis in Cloud Microservices Using Deep Learning},
year = {2020},
isbn = {978-3-030-76351-0},
publisher = {Springer-Verlag},
address = {Berlin, Heidelberg},
doi = {10.1007/978-3-030-76352-7_13},
booktitle = {Service-Oriented Computing  – ICSOC 2020 Workshops: AIOps, CFTIC, STRAPS, AI-PA, AI-IOTS, and Satellite Events, Dubai, United Arab Emirates, December 14–17, 2020, Proceedings},
pages = {85–96},
numpages = {12},
location = {Dubai, United Arab Emirates}
}

@ARTICLE{zhang2023diagfusion,
author={Zhang, Shenglin and Jin, Pengxiang and Lin, Zihan and Sun, Yongqian and Zhang, Bicheng and Xia, Sibo and Li, Zhengdan and Zhong, Zhenyu and Ma, Minghua and Jin, Wa and Zhang, Dai and Zhu, Zhenyu and Pei, Dan},
journal={IEEE Transactions on Services Computing}, 
title={Robust Failure Diagnosis of Microservice System Through Multimodal Data}, 
year={2023},
volume={16},
number={6},
pages={3851-3864},
doi={10.1109/TSC.2023.3290018}
}

@inproceedings{li2022dejavu,
author = {Li, Zeyan and Zhao, Nengwen and Li, Mingjie and Lu, Xianglin and Wang, Lixin and Chang, Dongdong and Nie, Xiaohui and Cao, Li and Zhang, Wenchi and Sui, Kaixin and Wang, Yanhua and Du, Xu and Duan, Guoqiang and Pei, Dan},
title = {Actionable and interpretable fault localization for recurring failures in online service systems},
year = {2022},
isbn = {9781450394130},
publisher = {Association for Computing Machinery},
address = {New York, NY, USA},
doi = {10.1145/3540250.3549092},
booktitle = {Proceedings of the 30th ACM Joint European Software Engineering Conference and Symposium on the Foundations of Software Engineering},
pages = {996–1008},
numpages = {13},
location = {Singapore, Singapore},
series = {ESEC/FSE 2022}
}

@article{brandon2020graph-based,
title = {Graph-based root cause analysis for service-oriented and microservice architectures},
journal = {Journal of Systems and Software},
volume = {159},
pages = {110432},
year = {2020},
issn = {0164-1212},
doi = {https://doi.org/10.1016/j.jss.2019.110432},
author = {Álvaro Brandón and Marc Solé and Alberto Huélamo and David Solans and María S. Pérez and Victor Muntés-Mulero},
}

@inproceedings{lee2023eadro,
author = {Lee, Cheryl and Yang, Tianyi and Chen, Zhuangbin and Su, Yuxin and Lyu, Michael R.},
title = {Eadro: An End-to-End Troubleshooting Framework for Microservices on Multi-Source Data},
year = {2023},
publisher = {IEEE Press},
doi = {10.1109/ICSE48619.2023.00150},
booktitle = {Proceedings of the 45th International Conference on Software Engineering},
pages = {1750–1762},
numpages = {13},
location = {Melbourne, Victoria, Australia},
series = {ICSE '23}
}

@inproceedings{yu2021microrank,
author = {Yu, Guangba and Chen, Pengfei and Chen, Hongyang and Guan, Zijie and Huang, Zicheng and Jing, Linxiao and Weng, Tianjun and Sun, Xinmeng and Li, Xiaoyun},
title = {MicroRank: End-to-End Latency Issue Localization with Extended Spectrum Analysis in Microservice Environments},
year = {2021},
publisher = {Association for Computing Machinery},
doi = {10.1145/3442381.3449905},
booktitle = {Proceedings of the Web Conference 2021},
pages = {3087–3098},
numpages = {12},
location = {Ljubljana, Slovenia},
series = {WWW '21}
}

@article{blondel2008louvain,
doi = {10.1088/1742-5468/2008/10/P10008},
year = {2008},
month = {oct},
publisher = {},
volume = {2008},
number = {10},
pages = {P10008},
author = {Blondel, Vincent D and Guillaume, Jean-Loup and Lambiotte, Renaud and Lefebvre, Etienne},
title = {Fast unfolding of communities in large networks},
journal = {Journal of Statistical Mechanics: Theory and Experiment},
}

@misc{kingma2017adamoptimization,
      title={Adam: A Method for Stochastic Optimization}, 
      author={Diederik P. Kingma and Jimmy Ba},
      year={2017},
      eprint={1412.6980},
      archivePrefix={arXiv},
      primaryClass={cs.LG},
      url={https://arxiv.org/abs/1412.6980}, 
}
}

\end{document}